# Cascade circuit architecture for RF-photonic frequency multiplication with minimum RF energy

MEHEDI HASAN[1,*] KARIN HINZER[1], TREVOR HALL,[1]

[1] *Photonic Technology Laboratory, Centre for Research in Photonics, University of Ottawa, 25 Templeton Street, Ottawa, K1N 6N5, ON, Canada*
*Corresponding author: mhasa067@uottawa.ca*

A general RF-photonic circuit design for implementing frequency multiplication is proposed. The circuit consists of $N$ cascaded differentially-driven Mach-Zehnder modulators biased at the minimum transmission point with a progressive RF phase shift of $\pi/N$ applied to each stage. The frequency multiplication factor obtained is $2N$. The novelty of the design is the reduced input RF energy required in comparison to the functionally equivalent parallel MZM configuration. Using transfer matrix method, a $N = 3$ architecture is modeled to obtain frequency sextupling. An industry standard simulation tool is used to verify the architectural concepts and analysis. The proposed design requires no optical or electrical filtering nor careful adjustment of the modulation index for correct operation. In addition, the overall intrinsic conversion efficiency of the $N = 3$ cascade circuit is improved by $\sim 5$ dB over a parallel MZM circuit. Finite MZM extinction and/or phase errors and power imbalances in the electric drive signals are also taken into consideration and their impact on overall performance investigated. The circuit can be integrated in any material platform that offers electro-optic modulators.

## I. INTRODUCTION

Conceptually the simplest means of generating tunable microwave carrier is the heterodyne beating on a high-speed photo detector of two optical carriers separated in frequency by the desired RF carrier frequency. The spectral purity of the microwave carrier depends on the phase correlation of the two optical carriers. External modulation of the laser is one of the stable ways to generate highly tunable spectrally pure microwave carriers with minimum system complexity.

To the best of the authors' knowledge using external modulation of a single laser source with a $LiNbO_3$ Mach Zehnder modulator (MZM) to perform RF carrier frequency doubling was first demonstrated in 1992 [1]. Since then numerous circuits of increasing sophistication with higher frequency multiplication factors have been proposed. For example, frequency quadrupling in [3-4], frequency six-tupling in [5-7], frequency octo-tupling in [2,7-11] and greater factors in [12-13] have been reported. The prior art has been summarised more comprehensively in Table1 of reference [2]. Very recently similar functionalities have been demonstrated using polarization optics [14-16]. The majority of the proposed circuits [1,3-5,9-16] use the same generalized Mach-Zehnder Interferometer (GMZI) architecture where $N$ parallel phase modulators (PM) are connected between a $1:N$ splitter and $N:1$ combiner. For even N, individual PMs are paired to form differentially driven MZMs. A lesser number of proposals have been made that make use of cascaded MZM architectures [2, 6-8]. Many of these proposed circuits have the demerit of requiring careful adjustment of the radio frequency (RF) drive levels and/or the use of optical filtering to suppress particular unwanted harmonics. These problems were resolved in reference [10] in which a photonic circuit architecture is presented for the first time that is capable of frequency octo-tupling without the aid of any optical or electrical filtering or careful adjustment of the modulation index. Subsequently, a two stage two-arm series-parallel architecture was reported [2] similarly free of filtering and precise modulation index adjustment that provided 3 dB greater RF conversion efficiency than the fully parallel implementation. The findings reported in reference [2] motivated the investigation reported herein of a single arm cascaded MZM architecture for frequency multiplication. To the best of author's knowledge, no such architecture is reported in the prior art. In this letter, a single stage cascaded generalized design principle is reported that can subsume all the designs in the prior art. At least in theory, the proposed design can generate $2N$ multiplication with significant RF advantage over a functionally equivalent parallel MZM architecture. where N is the number of MZM stages. As a proof of the concept, the general architecture is illustrated by an example of frequency sex-tupling by a three MZM stages in cascade ($N = 3$) and compared with a functionally equivalent parallel MZM structure [17]. The intrinsic conversion efficiency of the sex-tupling circuit is found to be 5 dB greater than the fully parallel architecture. The proposed photonic circuit can be integrated in any material platform that offers mature electro-optic phase modulators. The practical feasibility of cascading of multiple (15) MZM has already been demonstrated experimentally [18]. The critical part of the implementation of the design is to generate RF carriers with precise relative phase shift. However, the photonic generation of multiple phase controller with a wide tunable range ($-\pi$ to $\pi$) resolves the problem great authority [19].

## II. OPERATION PRINCIPLE

The general architecture is illustrated schematically in Fig 1. The circuit consists of $N$ cascaded MZM each biased at its MITP, each MZM driven differentially with RF drive signal having a $\pi/N$ progressive phase shift. The multiplication factor therby attained is $2N$. The advantage of a cascade of MZM is that the RF drive amplitude level required is reduced because all the end-to-end light paths traverse a cascade of $N$ phase modulators each driven by the same RF source albeit with a different phase. For simplicity, assume there is no transmission delay between the stages.

The architectures have a time-dependent transmission $f$ described by:

$$f(\omega t) = \prod_{p=0}^{N-1} g\left(\omega t + p\frac{\pi}{N}\right)$$
$$g(\omega t) = i\sin[m\cos(\omega t)]$$
(1)

where $n$ is the number of differential MZM stages, $p\pi/N$ is the RF-drive phase increment and $m$ is the modulation index, which is assumed the same for each stage. The time-dependent transmission $g$ is anti-periodic in $\omega t$ with period $\pi$. i.e. anti-periodicity with a specified period implies strict periodicity with twice the specified period.

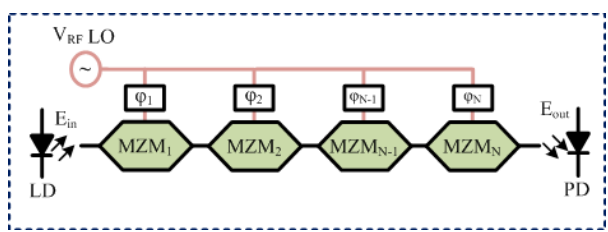

Fig. 1. Schematic diagram of the proposed frequency multiplier. LD: laser diode; PD: photo-diode; RF LO: radio frequency local oscillator.

.

$$g(\omega t + \pi) = -g(\omega t) \tag{2}$$

$$g(\omega t + 2\pi) = -g(\omega t + \pi) = g(\omega t) \tag{3}$$

for $\forall \omega t \in \mathbb{R}$. $g$ is therefore fully specified by its values over the base period $\omega t \in [0\ \pi]$. Within the base period $g$ possess a zero:

$$g(\omega t) = 0; \quad \Rightarrow \quad \omega t = \pi/2 \tag{4}$$

This is the only zero of $g$ provided $|m| < \pi$ and notably it has fixed position irrespective of the value of $m$. Additional zeros with modulation index dependent position are introduced if $|m| \geq \pi$. It follows that $f(\omega t)$ for $\omega t \in [-\pi, \pi]$ possess $2N$ (counting multiplicities) position independent zeros:

$$f(\omega t) = 0; \Rightarrow \omega t = \pm \pi/2 - p\frac{\pi}{N}\ mod(2\pi) \tag{5}$$

Where $p = 0, \dots, (n-1)$. Hence, there are $2N$ single zeros throughout $[-\pi, \pi)$ with a regular spacing of $\pi/N$. In addition, $f$ is anti-periodic with period $\pi/N$ since:

$$f\big(\omega t + (\pi/n)\big) = \prod_{p=0}^{n-1} g\left(\omega t + (p+1)\frac{\pi}{N}\right)$$
$$= -\prod_{p=0}^{n-1} g\left(\omega t + p\frac{\pi}{N}\right) = -f(\omega t) \tag{6}$$

A co-sinusoidal RF signal is thereby converted into a periodic amplitude modulation of the optical carrier with a $N-$fold submultiple of the period of the RF signal. Since $f$ is periodic over this smaller period it may be represented by the Fourier series:

$$f(\omega t) = \sum_{k=-\infty,\infty} c_k \exp(ikN\omega t) \tag{7}$$

where,

$$c_k = \frac{N}{2\pi} \int_{-\pi/N}^{\pi/N} f(\omega t) \exp(-ikN\omega t)\ d(\omega t) \tag{8}$$

Consequent all harmonics of the RF signal are suppressed (ideally perfectly) except for those that are a multiple of $N$. In addition, the carrier is suppressed due to the anti-periodicity. Table 1 summarizes the design conditions with required RF phase shift to achieve 4, 6 and 8 tupling with the respective RF advantages.

Table 1. Design parameters for specific operation with RF advantage.

| No. of MZM N | Unsuppressed harmonics | RF advantage (dB) | Progressive RF phase shift |
|---|---|---|---|
| 2 | $\{\dots -4, -2, +2, +4 \dots\}$ | 3 | $\pi/2$ |
| 3 | $\{\dots -9, -3, +3, +9 \dots\}$ | 6 | $\pi/3$ |
| 4 | $\{\dots -12, -4, +4, +12 \dots\}$ | 8 | $\pi/4$ |

As an example, a 3-stage cascaded MZM architecture is analysed in detail in this the letter. The time dependent transmission of $f(t)$ can be written as:

$$f(t) = i \sin(\emptyset_1)\, i \sin(\emptyset_2)\, i \sin(\emptyset_3) \tag{9}$$

where, $\emptyset_j = \pi v_j/v_\pi$, $j = 1, 2, 3$ is the optical phase shift induced the phase modulator with a half wave voltage $v_\pi$ within each arm of the MZM by the application of differentially driven RF signal with peak amplitude $v$. If the RF driving signal is $v_j = v \cos(\omega t + \varphi_{p+1})$ with $\varphi_{p+1} = p\pi/3$, Eq. (9) can be written as:

$$f(t) = -i \sin[z \cos(\omega t)] \sin[z \cos(\omega t + \pi/3)] \cdot \sin[z \cos(\omega t + 2\pi/3)] \tag{10}$$

where, $z = \pi v/v_\pi$.

Using trigonometric identities, Eq. (10) can be represented as:

$$f(t) = \frac{1}{4}i\{\sin(m \cos(\omega t - \pi/3)) - \sin(m \cos(\omega t)) + \sin(m \cos(\omega t + \pi/3))\} \tag{11}$$

where: $m = 2z$ is the modulation index.

Eq. (11) shows that for the same modulation index the required peak voltage $v$ is half that of a functionally equivalent single stage parallel architecture. Applying the Jacobi Anger expansion:

$$\sin(m \cos(\omega t + \theta)) = -2 \sum_{h=1}^{\infty} (-1)^h J_{2h-1}(m) \cos((\omega t + \theta)(2h-1)) \tag{12}$$

where $J_{2h-1}$ is the Bessel function of fisrt kind with order $h$. Developing right hand side of Eq. (11) using Eq. (12), yields:

$$f(t) = \frac{6}{4} i[J_3(m) \cos(3\omega t) - J_9(m) \cos(9\omega t) + \cdots] \tag{13}$$

or equivalently:

$$f(t) = \frac{3}{4}i \left[ J_3(m)e^{i3\omega t} + J_3(m)e^{-i3\omega t} - J_9(m)e^{i9\omega t} - J_9(m)e^{-i9\omega t} + \cdots \right] \tag{14}$$

It is seen from Eq. (14) that all the harmonics are supressed, except those for $h = 3(2k + 1)$, where $k$ is an integer. The equation also shows that the frequency six-tupling operation does not depends on the careful adjustment of modulation index rather it can be operated for a wide range of modulation index. If the modulation index is adjusted to suppress the 3rd harmonics, the proposed scheme can perform frequency 18-tupling.

## III. SIMULATION RESULTS

A VPI software simulation is used to verify the theoretical prediction proposed herein. A continuous wave distributed feedback (DFB) laser at a wavelength of 1550 nm with 10 mW of power is used as the optical input. A differentially driven MZM with a half wave voltage of 5 V is used from the VPI library. A PIN photodiode with a responsivity of 0.95 A/W is used to detect the frequency sextupled signal. A 10 GHz sinusoidal RF drive signal having an amplitude of 3.18 V is applied to each MZM with appropriate RF phase relation (i.e 0°, 60° and 120° for MZM$_1$, MZM$_2$ and MZM$_3$ respectively). Figure 2(a) and (b) shows the simulated optical spectrum of the dominant $\pm 3^{rd}$ harmonics and corresponding RF spectrum of the generated 60 GHz signal. The trace named "Ideal" in Fig 2(a, b) represents the optical and RF spectrum when the MZMs have infinite extinction and the RF drive signals have the correct amplitude and phase. In practice, the amplitude and phase of the RF drive signals may deviate from the ideal values. In addition to that, the extinction of the commercially available MZM is around 25-30 dB. The plot named "Non ideal$_1$" is obtained by setting the extinction of all MZMs to -30 dB. It is reasonable to assume that all MZMs on the same integrated circuit have similar extinction ratio as Yamazaki et al. obtained less than 0.5 dB loss variation among 4 MZMs [20]. It is seen from the optical spectrum that a side harmonics suppression ratio (SHSR) of 29 dB is achieved between the $\pm 3^{rd}$ harmonics and the optical carrier for an extinction ratio of 30 dB for all MZMs. The RF spectrum of the generated frequency sextupled signal is pristine. To emulate the effect of the amplitude variations or modulation index variation, the peak amplitude of the RF drive signals are set to 3.18V, 3.15V (approximately 3% error from the set value of 3.18 V) and 3.15V respectively for $MZM_1$, $MZM_2$ and $MZM_3$. Furthermore, the phases of the RF drive signal may not be precisely applied to the MZM as per the proposed design due to a path length mismatch between the RF drive signal and the MZM. It is found that the accuracy in length of the waveguide connected between RF drive signal and MZM is the most critical part of the design [10] in comparison to the power imbalances of the RF signal and/or low extinction ratio of the Mach Zehnder modulator.

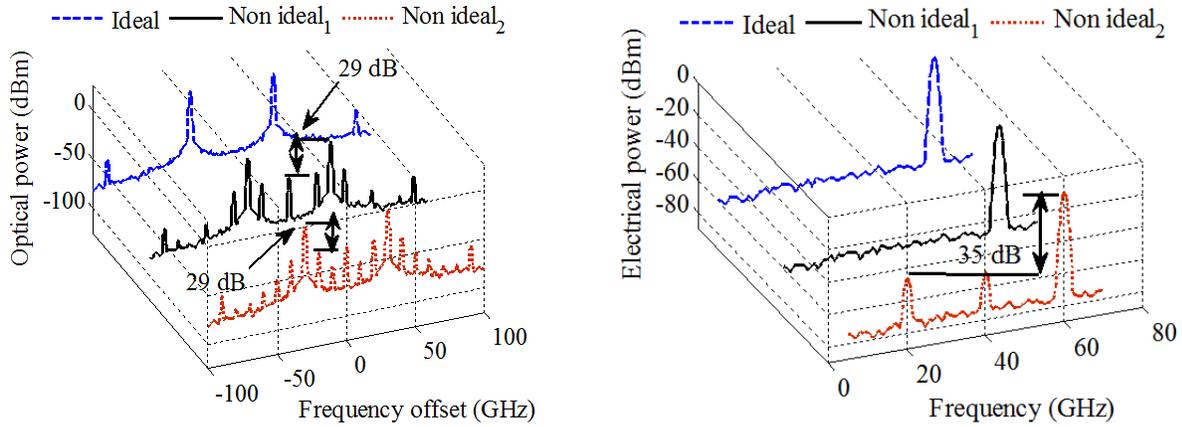

Fig. 2(a). Simulated optical spectrum analyzer. A resolution bandwidth of 1 GHz is used for the optical spectrum analyzer. (b). Simulated RF spectrum of the frequency sextupled signal. A resolution bandwidth of 1 GHz is used for the RF spectrum analyzer.

To asses the impact of the RF drive signal phase error on the proposed design, an incremental RF phase error is applied to the $MZM_2$ and the electrical side harmonic suppression ratio is recorded as shown in Fig 3. For example, the extinction ratio of all the MZMs is set to -30 dB first, secondly, the amplitude of the RF drive signals are set to 3.18 V, 3.15 V and 3.15 V respectively, then a one degree RF phase error (0°, 59° and 120° respectively) is applied to the $MZM_2$. As shown from the Fig. 3, an electrical SHSR of 26 dB is obtained. The upper inset shows the corresponding RF spectrum after photodetection. Still an electrical SHSR of 19 dB is achievable for a phase error of 2°. In practice, an electrical tunable delay line is required in the second and third stages to compensate the optical propagation delay caused by first and second stages. Hence, a fine tune of the delay line eliminates the phase error and thereby ensures a high electrical SHSR [21]. In addition, using modern digital synthesizer technology [22], not only 14-bit phase offset resolution but also 10 bit amplitude scaling resolution can be achieved.

The optical spectrum in Fig. 2(a) demonstrates that the proposed frequency 6-tupling operation does not require careful adjustment of the modulation index $m$ or optical filtering to suppress unwanted harmonics; rather it can be operated for a wide range of modulation index [2,10-11]. Hence, the advantages of driving the MZM at the peak of the particular Bessel function associated with the desired harmonics ($J_3$ in the example herein) to maximize the RF output can be taken. If it is permissible to adjust the modulation index to suppress the 3$^{rd}$ harmonics, the proposed design can be used to generate frequency 18-tupling. Figure 4 shows the comparison of RF power efficiency between the proposed architecture and a functionally equivalent 3-stage parallel MZM architecture. The $x$ axis of the plot shows the applied RF input power to each MZM in dBm, while the $y$ axis depicts the generated RF power for a load resistance of 50 $\Omega$. The half wave voltage of the MZM is considered as 5 V as representative of commercially available MZM. The plot depicts that for a same amount of RF input power (for example, 12 dBm) to each MZM, the RF output power of the proposed architecture is approximately 30 dB higher than parallel circuit. Alternatively, for the same output RF power, the proposed architecture requires ~5 dB less in input RF power than a 3-stage parallel MZM architectures when operating in the small modulation regime. The difference becomes 4 dB when the MZM is driven very hard because the cascade multiplication is about to die whereas the equivalent parallel stage architecture is remains in the small modulation index regime (e.g. 22 dBm RF power). The simulated results demonstrate that the proposed architecture can be used for an input RF power as low as 10 dBm.

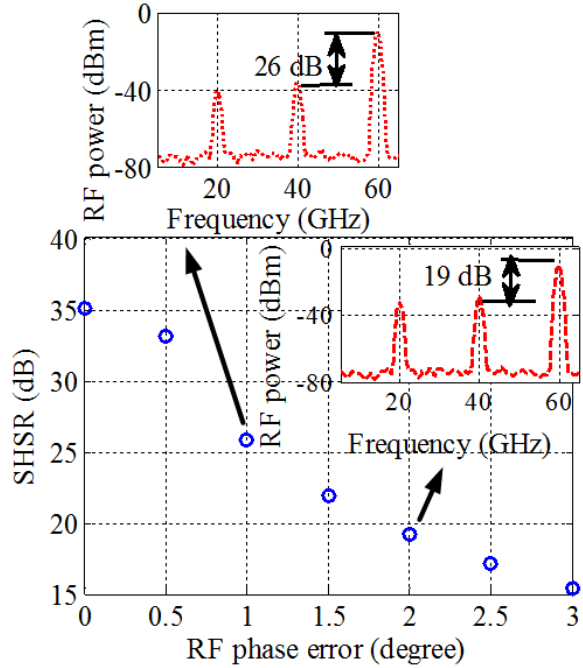

Fig 3. Simulated electrical SHSR for range of RF phase errors in the MZM$_2$. The inset shows the RF spectrum of the generated 60 GHz signal at the specified phase error.

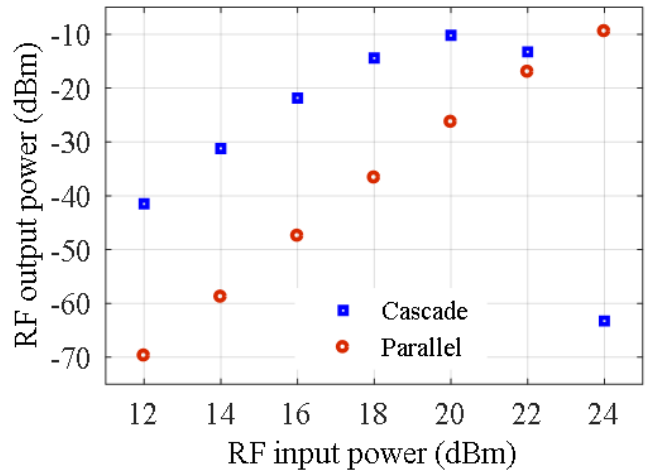

Fig 4. Comparison between proposed cascade architecture and functionally equivalent three stages parallel MZM configuration.

## IV. CONCLUSION

In summary, a generalized photonic circuit architecture is presented that subsumes and is more energy efficient than all the previous architectures that demonstrated photonic RF frequency multiplication techniques. As a proof of the concept, a 3-stage cascade MZM configuration is simulated and compared with functionally equivalent 3-stage parallel MZM configuration. A ~5 dB RF advantage is obtained for an like with like comparison. The proposed architecture can be integrated in any material platform that offers linear eletro-optic modulators. In addition to the commercially available $LiNbO_3$ technology, recent progress in phase modulator developments on silicon [23] or silicon hybrid [24], Si-$LiNbO_3$ hybridization (NTT) [20], and III-V [25] makes the implementation of the circuit practical.

**Acknowledgment.** Mehedi Hasan acknowledges the Natural Sciences and Engineering Research Council of Canada (NSERC) for their support through the Vanier Canada Graduate Scholarship program. Karin Hinzer is grateful to D & T Photonics for their support through the MITACS accelerate program. Trevor J. Hall is grateful to the University of Ottawa for their support through the University Research Chair program.